# What can we still learn from Brownian motion?

Qiuping A. Wang

ISMANS, 44 Ave. F.A., Bartholdi, 72000 Le Mans, France

LPEC, UMR 6087, Université du Maine, Ave. O. Messiaen, Le Mans, France

**Abstract**

Recent result of the numerical simulation of stochastic motion of conservative mechanical or weakly damped Brownian motion subject to conservative forces reveals that, in the case of Gaussian random forces, the path probability depends exponentially on Lagrangian action. This distribution implies a fundamental principle generalizing the least action principle of the Hamiltonian/Lagrangian mechanics and yields an extended formalism of mechanics for random dynamics. Within this theory, Liouville theorem of conservation of phase distribution breaks down. This opens a way to the Boltzmann $H$ theorem. We argue that the randomness is a crucial distinction between two kingdoms of Hamiltonian/Lagrangian mechanics: the stochastic dynamics and the regular one which is a special case of the first one for vanishing randomness. This distinction was missing in the criticisms of this theorem from Loschmidt, Poincaré and Zermelo.

1) Introduction

Brownian motion is a famous phenomenon observed by numerous scientists even before Brown in 1824[1] and an important member of the family of stochastic motion of mechanical systems. It is thoroughly investigated in experiment during longtime before the correct theoretical interpretation due to the pioneer work of Einstein[1]. Nowadays, this motion, well understood from statistical mechanics viewpoint, still raises wide interests in many fields of statistical mechanics, thermodynamics and many other multidisciplinary fields. It is also an experimental tool in the studies of fluctuation phenomena and free energy-work relations.

It is well known that, for regular motion obeying Newtonian mechanics of Hamiltonian systems, the path is unique between two given points in configuration space as well as in phase space[3] when the time period is specified. This is not the case for random dynamics in general. An example is Brownian motion. One remarkable characteristic of this motion is the non uniqueness of paths between two given points for given time period, which is illustrated in Figure 1 (left).

Although the path integral method has been an important tool for the study of Brownian motion in the past several decades[1], definitive answers to the following questions are still missing. What is the probability for a particle, moving between two points, to take a given path or to follow a thin bundle of paths among many others? What are the fluctuating variables of this probability? How to quantify the dynamical uncertainty in the path probability?

A point of view is that the path probability exponentially depends on the Hamiltonian or the time integral of Hamiltonian (Hamiltonian or Euclidean action) along the path, i.e., $A_H = \int_a^b (K+V)dt$ where $K$ is kinetic energy, $V$ potential energy. This assertion may find an example in the motion of a free Brownian particle implying a path probability proportional to $e^{-\gamma A_H}$ with $\gamma = 1/2mD$ where $m$ is the mass of the Brownian particle and $D$ the diffusion constant. Apart from the fact that this is the case of free motion ($V=0$ and $A_H = \int_a^b Kdt$), the Brownian motion is in general a damped or dissipative motion. Hence the involved energy is not only the kinetic energy. If the Brownian particle is subject to external conservative forces, the involved energy of the motion will include kinetic and potential energy as well as the energy dissipated to the environment. For dissipative fluctuation process, many works[5][6] argue that the probability of a path can be related to entropy production along that path. However, the explicit form of the path probability distribution is still understudied to our knowledge. Using the notion of entropy production in addition has its limit due to the random processes possibly far from equilibrium like many stochastic processes for which the notion of entropy is not well defined.

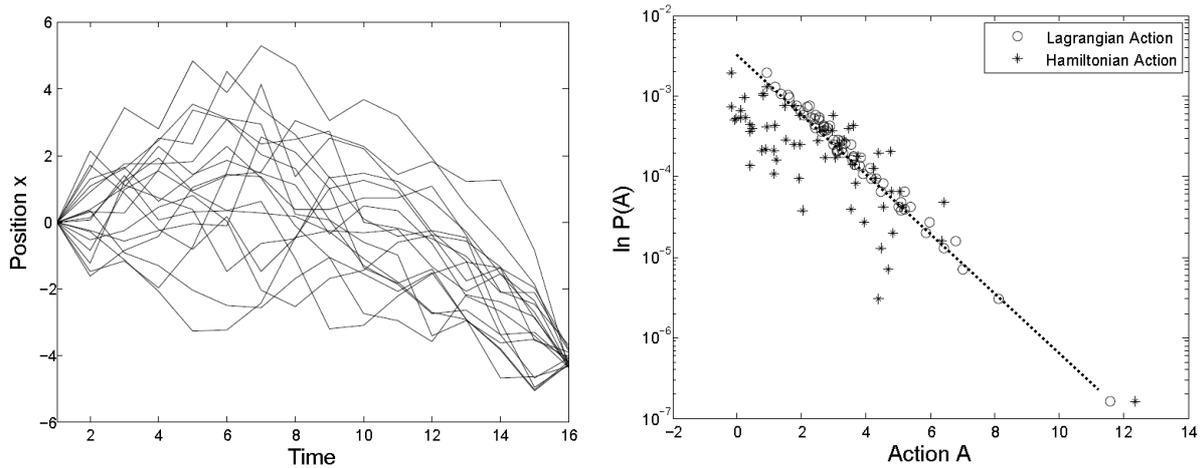

Figure 1: A result of numerical simulation of stochastic motion with $10^9$ particles subject to a constant force. The left panel shows the different sampled paths between two given points. The right panel shows the dependence of the probability distribution of paths $p(A)$ on the Lagrangian and Hamiltonian actions. The linearity of $p(A)$ with respect to the Lagrangian action (circles) and its independence from the Hamiltonian action (plus) are obvious. The most probable paths are the paths of least Lagrangian action (about 1 arbitrary unity) which is just the ballistic path of regular free fall. The straight line in the right panel is a guide for eye.

In our recent work[7], the probability distribution of paths of stochastic motion was studied by numerical simulation of a random dynamics of conservative Hamiltonian system or weakly damped Brownian motion subject to conservative force. No dissipation was considered since the aim of the work is to verify the prediction of a stochastic mechanics generalizing the Hamiltonian/Lagrangian mechanics to stochastic dynamics. Hence the system under consideration must be Hamiltonian system or, more precisely, conservative Hamiltonian system. Nevertheless, it has been verified that the results are correct as well for weakly damped stochastic (or Brownian-like) motion. The technical details about the simulation model can be found in reference [7]. In this simulation, the path probability was determined by counting the number of particles passing by each sampled path between two given points. The Lagrangian and the Hamiltonian actions are calculated for paths. Figure 1 shows an example of the results. The exponential dependence of the path probability on the Lagrangian action is obvious from this result. From now on, we will drop the index of action. The Lagrangian action is denoted by $A$ or $A_k(a,b)$ for a given path $k$ between the points $a$ and $b$. The result in the right panel of Figure 1 can be expressed by

$$p_k(a,b) = \frac{1}{Z_{ab}} e^{-\gamma A_k(a,b)}. \qquad (1)$$

where $\gamma$ is the slope of the straight doted line in the right panel and $Z_{ab}$ is the partition function given by $Z_{ab} = \sum_k e^{-\gamma A_k(a,b)}$ summed over all the possible paths between *a* and *b*. In general, this summation can be carried out by path integral[4].

In this paper, we present a theoretical analysis of this result and some consequences. This numerical result verifies the prediction of a stochastic action principle (SAP). In the framework of the Hamiltonian/Lagrangian mechanics based on this SAP, the usual least action principle applies only in the special case of vanishing randomness, i.e., the case of the regular Hamiltonian/Lagrangian mechanics. An important theoretical consequence of this is the modification of Liouville theorem of conservation of phase volume. Other consequences relative to thermostatistics are also presented.

## 2) Stochastic least action principle

The probabilistic uncertainty of the distribution Eq.(1) can be measured by the Shannon information formula $S_{ab} = -\sum_k p_k(a,b)\ln p_k(a,b)$ called path entropy. It is easy to calculate

$$S_{ab} = \ln Z_{ab} + \gamma \overline{A}_{ab}. \qquad (2)$$

On the other hand, the distribution of Eq.(1) maximizes the path entropy. In other words, the vanishing variation $\delta(S_{ab} - \gamma \overline{A}_{ab}) = 0$ yield Eq.(1). Considering that $\delta \ln Z_{ab} = \overline{\delta A}$, we get

$$\overline{\delta A} = 0 \qquad (3)$$

where $\delta A$ is a variation of the Lagrangian action and the average is over all possible paths between *a* and *b*. This formula was called stochastic least action principle (SAP) proposed in our previous work as an axiom [8]-[11], while here it is derived from the above mentioned numerical result.

## 3) Hamiltonian mechanics revisited

It has been shown that the exponential probability distribution of action satisfied the Fokker-Planck equation for normal diffusion in the same way as the Feymann factor of quantum propagator $P_k \propto e^{-iA_k/\hbar}$ satisfies the Schrödinger equation[4]. Here we will show a generalized formalism of Hamiltonian mechanics with equations which we will use for the discussion of Liouville theorem and Boltzmann *H* theorem.

### a) Euler-Lagrange equations

A meaning of the SAP Eq.(3) is that for any particular path *k*, there is no necessarily $\delta A_k = 0$. In general we have

$$\delta A_k = \int_a^b \left[ \frac{\partial}{\partial t}\left(\frac{\partial L_k}{\partial \dot{x}}\right) - \frac{\partial L_k}{\partial x} \right] \delta x\, dt \geq \text{ (or } \leq \text{) } 0 \qquad (4)$$

where $\delta x$ is an arbitrary variation of *x* which is zero at *a* and *b*. For $\delta A_k \geq 0$ (or $\leq 0$), we get

$$\frac{\partial}{\partial t}\left(\frac{\partial L_k}{\partial \dot{x}}\right) - \frac{\partial L_k}{\partial x} \geq \text{ (or } \leq \text{) } 0 \qquad (5)$$

which can be proved by contradiction as follows. Suppose $\int_a^b f(t)\delta x\, dt \geq 0$ and $f(t) \leq c \leq 0$ during a small period of time $\Delta t$ somewhere between *a* and *b*. Since $\delta x$ is arbitrary, let it be zero outside $\Delta t$ and a positive constant within $\Delta t$. We clearly have $\int_a^b f(t)\delta x dt \leq c\delta x \leq 0$, which contradicts our starting assumption. This proves Eq.(5).

The Legendre transformation $H_k = P_k \dot{x} - L_k$ along a path $k$ implies the momentum given by $P_k = \frac{\partial L_k}{\partial \dot{x}}$ which can be put into Eq.(5) to have

$$\dot{P}_k \geq (\text{or} \leq) \frac{\partial L_k}{\partial x} \qquad (6)$$

for $\delta A_k \geq (or \leq) 0$.

However, from the path average of Eq.(4) and the SAP $\overline{\delta A} = 0$, we straightforwardly write

$$\overline{\delta A} = \sum_k p_k \int_a^b \left[\frac{\partial}{\partial t}\left(\frac{\partial L_k}{\partial \dot{x}}\right) - \frac{\partial L_k}{\partial x}\right]\delta x dt = \int_a^b \left[\overline{\frac{\partial}{\partial t}\left(\frac{\partial L_k}{\partial \dot{x}}\right)} - \overline{\frac{\partial L_k}{\partial x}}\right]\delta x dt = 0 \qquad (7)$$

which implies

$$\overline{\frac{\partial}{\partial t}\left(\frac{\partial L_k}{\partial \dot{x}}\right)} - \overline{\frac{\partial L_k}{\partial x}} = 0. \qquad (8)$$

This is the Euler-Lagrange equation of the random dynamics. We have equivalently

$$\overline{\dot{P}} = \overline{\frac{\partial L}{\partial x}}. \qquad (9)$$

where $\overline{\dot{P}} = \sum_k p_k \dot{P}_k$ and $L = \sum_k p_k L_k$.

*b) Hamiltonian equations*

From Legendre transformation, we can have $\frac{\partial L_k}{\partial x} = -\frac{\partial H_k}{\partial x}$ and the following Hamiltonian equations

$$\dot{x}_k = \frac{\partial H_k}{\partial P_k} \text{ and } \dot{P}_k \geq (\text{or} \leq) -\frac{\partial H_k}{\partial x} \qquad (10)$$

For a path $k$ along which $\delta A_k \geq (or \leq) 0$.

Naturally, Eq.(8) means

$$\overline{\dot{P}} = \overline{\frac{\partial H}{\partial x}}. \qquad (11)$$

with the average Hamiltonian $H = \sum_k p_k H_k$.

## 4) Liouville theorem

The Liouville theorem is often involved in the discussions relative to thermodynamic entropy in statistical mechanics. We give an outline below followed by an analysis of the theorem within the present formulation of (probabilistic) Hamiltonian mechanics.

We look at the time change of phase point density $\rho(x, P, t)$ in a, say, 2-dimensional phase space $\Gamma$ when the system of interest moves on the geodesic, i.e., the path of least action[3]. $\rho(x, P, t)$ can also be considered as the density of systems of an ensemble of a large number of systems moving in phase space. The time evolution neither creates nor destroys state points or systems, hence the law of state conservation in the phase space is

$$\frac{\partial \rho}{\partial t} + \frac{\partial (\dot{x}\rho)}{\partial x} + \frac{\partial (\dot{P}\rho)}{\partial P} = 0 \tag{12}$$

which means

$$\frac{d\rho}{dt} = \frac{\partial \rho}{\partial t} + \frac{\partial \rho}{\partial x}\dot{x} + \frac{\partial \rho}{\partial P}\dot{P} = -\left(\frac{\partial \dot{x}}{\partial x} + \frac{\partial \dot{P}}{\partial P}\right)\rho. \tag{13}$$

For the least action path satisfying Hamiltonian equations[3], the right hand side of the above equation is zero, leading to the Liouville theorem

$$\frac{d\rho}{dt} = 0, \tag{14}$$

i.e., the state density in phase space is a constant of motion. The phase volume $\Omega$ available to the system can be calculated by $\Omega = \int_\Gamma \frac{1}{\rho} dn$ where $dn$ is the number of phase point in an elementary volume $d\Gamma$ at some point in phase space. The time evolution of the phase volume $\Omega$ accessible to the system is then given by

$$\frac{d\Omega}{dt} = \frac{d}{dt}\int_\Gamma \frac{1}{\rho} dn = -\int_\Gamma \frac{1}{\rho^2}\frac{d\rho}{dt} dn = 0 \tag{15}$$

meaning that this phase volume is a constant of motion.

The second law of thermodynamics states that the entropy of an isolated system increases or remains constant in time. But the Liouville theorem implies that if the motion of the system obeys the fundamental laws of mechanics, the Boltzmann entropy defined by $S = \ln\Omega$ must be constant in time. On the other hand, the probability distribution of states $p(x,P)$ in phase space is proportional to $\rho(x,P)$, meaning that an entropy $S(p)$, as a functional of $p(x,P)$, must be constant in time, which is in contradiction with the second law.

What is then the Liouville theorem in the context of random dynamics? From Eq.(13), we have $\left.\frac{d\rho}{dt}\right|_k = -\left(\frac{\partial \dot{x}_k}{\partial x} + \frac{\partial \dot{P}_k}{\partial P_k}\right)\rho$ along a path $k$. Let us write the second equation of the Eqs.(10) as $\dot{P}_k = -\frac{\partial H_k}{\partial x} + R_k$ where $R_k \geq (or \leq) 0$ for $\delta A_k \geq (or \leq) 0$ is the random force causing the deviation from Newtonian laws. The average Newtonian law $\overline{\dot{P}} = \overline{\frac{\partial H}{\partial x}}$ yields $\sum_k p_k R_k = 0$ for any moment of the process. We then have

$$\left.\frac{d\rho}{dt}\right|_k = -\frac{\partial R_k}{\partial P_k}\rho. \tag{16}$$

and

$$\frac{d\rho}{dt} = \sum_k p_k \left.\frac{d\rho}{dt}\right|_k = -\overline{\frac{\partial R_k}{\partial P_k}}\rho. \tag{17}$$

where $\overline{\frac{\partial R_k}{\partial P_k}} = \sum_k p_k \frac{\partial R_k}{\partial P_k}$ is an average over all the possible paths. The solution of this equation is

$$\rho(t) = \rho(t_0)\exp[\zeta(t,t_0)] \qquad (18)$$

with the function $\zeta(t,t_0) = -\int_{t_0}^{t} \overline{\frac{\partial R_k}{\partial P_k}} dt$.

Now let us see an application to the case of exponential path probability. The relationship $\sum_k p_k R_k = 0$ implies $\overline{\frac{\partial R_k}{\partial P_k}} = -\sum_k R_k \frac{\partial p_k}{\partial P_k}$. By using the exponential distribution of action, it can be proved that $\frac{\partial p_k}{\partial P_k} = -\gamma p_k dx_k$, where $dx_k$ is a displacement of the motion at time $t$ along the path $k$. Hence $\overline{\frac{\partial R_k}{\partial P_k}} = \gamma \sum_k p_k R_k dx_k = \gamma \overline{\delta W_R(t)}$ where $\overline{\delta W_R(t)} = \sum_k p_k dW_{Rk}$ is the average of the random work $dW_{Rk} = R_k dx_k$ performed by the random forces over the random displacement $dx_k$ along the path $k$ at time $t$. Finally, $\zeta(t,t_0) = -\gamma \int_{t_0}^{t} \overline{\delta W_R(t)} dt = -\gamma W_R(t,t_0)$ here $W_R(t,t_0) = \int_{t_0}^{t} \overline{\delta W_R(t)} dt$ is the cumulate average work of random force performed from $t_0$ and $t$. We have

$$\rho(t) = \rho(t_0)\exp[-\gamma W_R(t,t_0)] \qquad (19)$$

meaning that the state density decreases (increases) and the phase volume increases (decreases) whenever the cumulate random work $W_R(t,t_0) > 0$ ($< 0$). $\rho(t)$ is constant only when there is no work of random forces.

If the average random work $\overline{\delta W_R(t)}$ does not depend on time, we have $W_R(t,t_0) = \overline{\delta W_R}(t-t_0)$ and

$$\rho(t) = \rho(0)\exp[-\lambda t] \qquad (20)$$

where $\lambda = \gamma \overline{\delta W_R}$ is a Lyapunov-like exponent characterizing the variation of the distances between the state points.

5) Boltzmann *H* theorem

The Boltzmann H function can be defined in discrete coarse graining way in phase space by

$$H(t) = \sum_{x,P} p(x,P,t) \ln p(x,P,t) \qquad (21)$$

where the sum is over all the accessible coarse grained phase domain and $p(x,P,t) \propto \rho(x,P,t) d\Omega$ is the probability that a system is found in a phase cell of volume $d\Omega$ situated at the point $(x,P)$ at time $t$.

From Eq.(13), we get

$$p(x,P,t) = p(x_0,P_0)\exp[-\gamma W_R(t,t_0)]. \qquad (22)$$

It is straightforward to see that the variation of $H$ function from $t_0$ to $t$ is given by $\Delta H = H(t) - H(t_0) = -\gamma W_R(t,t_0)$. In order to yield Boltzmann $H$ theorem $\Delta H \leq 0$, it is necessary to prove that the cumulate work $\gamma W_R(t,t_0)$ is positive in a general way.

Here we only provide a case of $\mathcal{W}_R(t,t_0) \geq 0$ with ideal gas. In this isolated gas each molecule undergoes stochastic motion with large fluctuation. The movement of the ensemble of the molecules is also stochastic due to internal thermal fluctuation.

Let us consider the free expansion of an ideal gas between two equilibrium states at time $t_1$ and $t_2$ with volume $V_1$ and $V_2$, respectively. Let $p(x_1,P_1,t_1)$ and $p(x_2,P_2,t_2)$ be the two equilibrium distributions. We use the entropy expression

$$S(t) = \sum_{x,P} p(x,P,t) \ln p(x,P,t) \quad (23)$$

For the two equilibrium states. It is straightforward to get $\Delta S = S_2 - S_1 = N \ln \frac{V_2}{V_1} = -\zeta(t,t_0) = \mathcal{W}_R(t,t_0)$ (Boltzmann constant $k_B=1$). Hence the variation of $H$ function is $\Delta H = H_2 - H_1 = -\Delta S = -N \ln \frac{V_2}{V_1}$, meaning that $\mathcal{W}_R(t,t_0) = N \ln \frac{V_2}{V_1} \geq 0$. This is a proof for the validity of $H$ theorem in the special case of ideal gas expansion between two equilibrium states. For the motion between any two states, the property $\mathcal{W}_R(t,t_0) \geq 0$ needs general proofs which will be given in our later work.

6) Concluding remarks

Recent results of numerical simulation of stochastic motion of Hamiltonian system subject to conservative forces showed a path probability depending exponentially on Lagrangian action. Based on this, we presented here some theoretical aspects related to a variational principle generalizing the least action principle of the Hamiltonian/Lagrangian mechanics. Within the framework of the mechanics theory derived from this generalized least action principle, we provided a modification of Liouville theorem for this family of random dynamics. Since the phase space distribution function is no more constant in time, $H$ function can have time evolution. It can be proved for the time being that this is a decreasing evolution between two equilibrium states. More general proof for the evolution between any states is in progress.

One of the theoretical consequences is that the criticisms of $H$ theorem from Loschmidt, Poincaré and Zermelo[12]-[14] on the basis of the Liouville theorem lose their cornerstone. The crucial flaw in these criticisms is the missing distinction between regular mechanical motion and stochastic dynamics which is ubiquitous in thermodynamic systems.

In fact, in order to use Liouville theorem to criticize $H$ theorem, one should take it for granted that the state density in Liouville theorem is proportional to the probability distribution of states. Yet the Liouville theorem holds only for regular dynamics in which each trajectory can be a priori traced in time with certainty, each state has its unique moment of time to be visited by the system, and the frequency of visit of any phase (state) volume can be predicted exactly. Therefore *in principle* there is no place for probability and entropy or other uncertainty in regular dynamics when Liouville theorem holds. If ever it is necessary *in practice* to use a distribution function of states in phase space due to very large number of degrees of freedom overpassing our capacity of observation and description, this distribution function does not mean an intrinsic and objective probabilistic property of the dynamics. This pragmatic and subjective "probability" distribution has nothing to do with the objective probability distribution which should be used in $H$ function and in the concomitant entropy of the stochastic motions of thermodynamics. Objective probability exists only when the dynamics is intrinsically probabilistic and random, meaning that not only an observer does not know where a system is going at next time step, the system itself also "ignores" it. From this point of view, the very first idea from Boltzmann, Clausius, Hertz and Helmholtz[15] to relate the second law of thermodynamics to the Hamiltonian/Lagrangian mechanics is impossible.

...x